\documentclass[aps,preprintnumbers,superscriptaddress,showpacs]{revtex4}
\usepackage{epsfig}
\usepackage{psfrag}
\usepackage{amsfonts}
\usepackage{graphicx}
\usepackage{dcolumn}
\usepackage{bm}
\usepackage{amsmath}

\begin{document}

\title{Holographic light quark jet quenching in flavor resolved QCD plasmas}

\author{Hu-wei Zhu}
\email{zhuhuwei@cug.edu.cn} \affiliation{School of Mathematics and
Physics, China University of Geosciences (Wuhan), Wuhan 430074,
China}

\author{Ke Ma}
\email{make@cug.edu.cn} \affiliation{School of Mathematics and
Physics, China University of Geosciences (Wuhan), Wuhan 430074,
China}

\author{Zi-qiang Zhang}
\email{zhangzq@cug.edu.cn} \affiliation{School of Mathematics and
Physics, China University of Geosciences (Wuhan), Wuhan 430074,
China}

\begin{abstract}
We study light quark energy loss in a quark-gluon plasma using the
holographic falling-string framework on a
machine-learning-calibrated Einstein-Maxwell-dilaton background
for three flavor configurations: pure glue, two-flavor QCD, and
full QCD with strangeness. Stopping distances are computed from
null geodesics as functions of reduced temperature $T/T_c$ and
baryon chemical potential $\mu$. For $T/T_c>1.03$, the flavor
ordering of the integrated stopping distance stabilizes to
$N_f=2+1>N_f=2>N_f=0$, corresponding to the quenching hierarchy
$N_f=0>N_f=2>N_f=2+1$. This agrees with the flavor ordering
obtained from the shooting-string $dE/dx$ when both are expressed
in the reduced-temperature framework, providing a cross-check
between local and integrated observables within the same
holographic background. Near $T/T_c\simeq1$, the behavior is more
complex due to the interplay between the $a$ and $b$ terms in
$A(z)$ and the different $T_c$ values of the three systems. Finite
baryon chemical potential monotonically shortens stopping
distances and enhances quenching, with enhanced sensitivity near
the QCD critical endpoint, consistent with expectations from RHIC
beam energy scan measurements. We discuss the geometric origin of
the high-$T/T_c$ flavor ordering and possible extensions to
heavy-ion transport simulations.
\end{abstract}

 \maketitle

\section{Introduction}

Ultrarelativistic heavy ion collisions at RHIC and the LHC have
established that the quark-gluon plasma (QGP) created in central
collisions behaves as a nearly perfect liquid, characterized by an
extremely small shear viscosity to entropy ratio $\eta/s$
\cite{Shuryak:2004,Kovtun:2005,Adams:2005}. This value lies well
below perturbative QCD (pQCD) predictions, indicating that the
experimentally realized QGP resides in a strongly coupled regime
where standard perturbative methods based on weak coupling
expansions cease to be reliable \cite{Blaizot:2015}. Lattice QCD
provides first principles access to equilibrium thermodynamics but
suffers from the sign problem at finite baryon density and cannot
directly simulate real time, non-equilibrium processes such as
hard parton propagation and medium induced energy loss
\cite{Aarts:2016}. The gauge/gravity duality, or holography,
offers a powerful alternative nonperturbative tool: it maps
strongly coupled four dimensional gauge theories onto weakly
coupled five dimensional gravitational spacetimes, circumventing
the limitations of both pQCD and lattice QCD
\cite{Maldacena:1998,Gubser:1998,Witten:1998}.

Over the past three decades, holography has been widely used to
characterize transport coefficients and thermodynamic properties
of the QGP, with jet quenching emerging as one of its most
promising phenomenological applications
\cite{CasalderreySolana:2012,DeWolfe:2014}. Jet quenching
describes the energy degradation experienced by high transverse
momentum ($p_T$) partons as they traverse the hot, deconfined QGP
medium. Energetic light and heavy quarks, produced in early hard
scatterings, interact with the medium through elastic collisions
and induced gluon radiation, gradually losing energy and softening
the final state jet spectrum \cite{Wang:1992,Majumder:2011}.
Experimentally, the nuclear modification factor $R_{AA}$
quantifies the suppression of high $p_T$ hadron yields in
nucleus-nucleus collisions relative to proton-proton reference
data, while elliptic flow $v_2$ encodes the collective geometric
expansion of the QGP fireball \cite{Qin:2015}.

A particularly important aspect of QGP phenomenology is the role
of dynamical quark flavors and finite baryon density. RHIC beam
energy scans have revealed nonmonotonic behavior of various
observables as the collision energy varies, suggesting the
possible presence of a critical endpoint (CEP) in the QCD phase
diagram \cite{Arsene:2005}. Meanwhile, LHC measurements have
demonstrated that strange hadron production and suppression
patterns carry valuable information about the flavor structure of
the QGP \cite{ATLAS_strange_UE_2024}. Furthermore, precision
measurements of jet substructure in heavy ion collisions have
provided new insights into the mechanism of in-medium parton
energy loss and its dependence on the jet internal structure
\cite{ATLAS_jet_substructure_2025}. These observations indicate
that dynamical strange quarks and finite baryon density
substantially modify parton-medium coupling, effects that cannot
be captured by minimal conformal $\mathcal{N}=4$ super-Yang-Mills
(SYM) holographic models which lack dynamical quark flavors and
baryonic charge.

The need to go beyond the conformal SYM paradigm has motivated the
development of more realistic holographic QCD models
\cite{JBJ,MKD,TSS,SH2,HJ,OA1,AKE,JPF,KGN,UGE,D1,D2,Chen:2024} that
incorporate nonconformal features of QCD, including a running
coupling, confinement-deconfinement transition, and dynamical
quark flavors. A major breakthrough in this direction was achieved
by Chen et al. \cite{Chen:2024}, who employed machine learning
techniques to construct an Einstein-Maxwell-dilaton (EMD)
holographic framework whose thermodynamics are precisely
calibrated to lattice QCD data. This framework systematically
incorporates the effects of dynamical light and strange quarks
through flavor dependent bulk metric parameters, yielding distinct
zero density crossover temperatures and critical endpoint
locations for different flavor configurations. Since its
construction, this framework has been applied to a variety of QCD
observables, including the equation of state, transport
coefficients, and jet-quenching related quantities
\cite{m1,m2,m3,m4,m5,m6,m7,m8,m9,m10,m11}.

Within holographic QCD, two distinct open string configurations
describe light quark energy loss. The shooting string setup places
string endpoints carrying finite momentum near the black brane
horizon; these endpoints then propagate outward toward the AdS
boundary \cite{ss1,ss2}. Energy continuously flows from the
energetic endpoint into the trailing string, directly yielding the
local instantaneous energy loss rate $dE/dx$ at each point along
the trajectory. A recent comprehensive calculation
\cite{Zhang:2025} implemented shooting strings on the machine
learning optimized, flavor dependent EMD holographic background.
That work established that additional $u/d$ and strange $s$ quarks
increase the magnitude of $|dE/dx|$, while energy loss weakens as
the system approaches the QCD CEP and grows monotonically as
$(T,\mu)$ moves away from criticality.

In contrast, the falling string formalism describes massless
boundary gauge wave packets originating near the AdS boundary,
dual to high energy light quarks injected into the QGP
\cite{ss,pm,pm1,pa,pa1}. Under the semiclassical
Wentzel-Kramers-Brillouin (WKB) approximation, these bulk
excitations follow null geodesics. Full thermalization of the
boundary jet occurs once the string endpoint falls entirely
through the black brane horizon. The maximal longitudinal distance
traveled by the quark before horizon absorption defines the
thermalization stopping distance $x_{\mathrm{stop}}$, an
integrated observable that encodes the total cumulative energy
loss over the entire in-medium path. Earlier falling string
studies were mostly restricted to the conformal $\mathcal{N}=4$
SYM spacetime, omitting realistic medium effects constrained by
lattice QCD
\cite{YD,Zhu:2019,Zhang:2023,Zhang:2019a,Zhang:2019b,Zhu:2021}.

The machine learning calibrated EMD framework now permits a
systematic evaluation of falling string stopping distances on a
realistic, flavor dependent holographic background. Such an
investigation has several useful purposes. First, it offers a
direct quantitative cross-check against the shooting string
$dE/dx$ results obtained on the identical geometry, testing the
consistency of flavor dependent quenching trends across two
complementary string formalism. Second, the stopping distance, as
an integrated observable, connects more directly to experimentally
measured jet suppression patterns than the local energy loss rate.
Third, a unified analysis of both differential and integrated
observables within the same holographic framework would strengthen
the theoretical foundation for phenomenological applications to
heavy ion collision data.

Systematic evaluations of falling string stopping distances on the
lattice calibrated EMD background are largely absent from the
existing literature. This leaves a gap in quantitative
cross-validation between the two string approaches and obscures
the relationship between flavor modulated holographic jet
quenching signatures and the experimental measurements collected
at RHIC and the LHC. The present study addresses this by carrying
out a unified holographic analysis that links complementary string
observables to heavy ion experimental signatures.

Specifically, we numerically compute light quark stopping
distances from falling string null geodesics on the same lattice
fitted, flavor dependent EMD spacetime used in the shooting string
study of \cite{Zhang:2025}. We systematically compare temperature,
flavor content, and baryon density trends against the shooting
string energy loss data, enabling a comparison between local and
integrated observables within the same holographic framework. We
also establish qualitative connections between our holographic
quenching predictions and key experimental observables, including
critical endpoint anomalies observed in beam energy scans and
strange hadron suppression patterns. These results provide a
theoretical foundation for future quantitative fits of holographic
transport models to collision data.

The remainder of this paper is organized as follows. In
Sec.~\ref{sec:background}, we introduce the five dimensional EMD
gravity action, metric ansatz, analytic bulk solutions, and
machine learning fitted parameters from \cite{Chen:2024}, ensuring
full geometric consistency with the companion shooting string
study. In Sec.~\ref{sec:fallingstring}, we review the standard WKB
derivation of the falling string formalism for evaluating stopping
distances, following \cite{pa,pa1}, and describe the construction
of the conformal $\mathcal{N}=4$ SYM reference baseline used for
normalization. Our numerical results are presented and analyzed in
Sec.~\ref{sec:results}, covering zero density flavor ordering,
finite baryon potential sensitivity near the CEP, and direct
comparison with shooting string results. We then elaborate on the
competing bulk geometric effects responsible for the high
temperature flavor crossover in Sec.~\ref{sec:discussion}, a
unique prediction of our flavor resolved EMD setup that is absent
in most earlier holographic jet studies. Finally,
Sec.~\ref{sec:conclusion} summarizes our findings, discusses
phenomenological implementation of holographic stopping distances
within heavy ion transport codes, and proposes multiple directions
for follow-up research.

\section{Background geometry}
\label{sec:background}

The holographic modeling of QCD thermodynamics requires a
gravitational framework that can reproduce key features of the
strongly coupled plasma, including the nonconformal equation of
state, the confinement-deconfinement transition, and the response
to finite baryon density. The EMD gravity framework has emerged as
one of the most successful holographic approaches in this regard,
as it provides sufficient flexibility to encode the essential
physics while remaining computationally tractable.

We adopt the same five dimensional EMD gravity framework employed
for shooting string calculations in \cite{Zhang:2025}, with metric
parameters optimized via machine learning to reproduce lattice QCD
thermodynamics for pure gluon, two flavor, and $2+1$ flavor QCD
media. The full gravitational action in the Einstein frame reads
\cite{Chen:2024}
\begin{equation}
S = \int \frac{d^5 x}{16\pi G_5} \sqrt{-\det(g_{\mu\nu})} \left[R
- \frac{f(\phi)}{4} F_{\mu\nu}F^{\mu\nu} - \frac{1}{2}
\partial_\mu \phi \partial^\mu \phi - V(\phi)\right],
\end{equation}
where $G_5$ is the five dimensional Newton constant, $R$ the Ricci
scalar, $F_{\mu\nu}=\partial_\mu A_\nu-\partial_\nu A_\mu$ the
U(1) Maxwell field strength dual to the boundary baryon number
current, $\phi$ the dilaton field, $f(\phi)$ the dilaton dependent
gauge kinetic function, and $V(\phi)$ the dilaton potential.
Varying this action yields coupled Einstein-dilaton-Maxwell field
equations, whose closed analytic solutions can be obtained once a
warped black brane metric ansatz is specified.

The standard static, homogeneous, isotropic black brane metric
widely adopted in finite density dilaton holographic models takes
the form
\begin{equation}
ds^2 = \frac{L^2 e^{2A(z)}}{z^2} \left[ -g(z) dt^2 + d\vec{x}^2 +
\frac{dz^2}{g(z)} \right],
\end{equation}
where $L$ is the AdS curvature radius and $z$ labels the fifth
dimensional radial coordinate. The physical four dimensional QGP
boundary lies at $z\to0$, and the black brane horizon is located
at finite $z=z_t$, where $g(z_t)=0$. The warp factor $A(z)$
encodes dilaton modulated nonconformal QCD corrections, while the
blackening factor $g(z)$ controls the bulk causal structure and is
subject to the horizon regularity condition $g(z_t)=0$. The
temporal bulk gauge field $A_t(z)$ encodes the boundary baryon
chemical potential $\mu$, satisfying $A_t(0)=\mu$ and
$A_t(z_t)=0$.

Following the machine learning fitting procedure detailed in
\cite{Chen:2024}, the closed analytic expressions for $g(z)$,
$\phi'(z)$, $A_t(z)$, and $V(z)$ are:
\begin{align}
g(z) &= 1 - \frac{1}{\displaystyle\int_{0}^{z_t}dx \, x^3 e^{-3A(x)}} \left[ \int_{0}^{z} dx \, x^3 e^{-3A(x)} + \frac{2c\mu^2 e^k}{(1 - e^{-c z_t^2})^2} \det \mathcal{G} \right], \label{blackening factor} \\
\phi'(z) &= \sqrt{6\left(A'^2 - A'' - 2A'/z\right)}, \\
A_t(z) &= \mu \frac{e^{-c z^2} - e^{-c z_t^2}}{1 - e^{-c z_t^2}}, \\
V(z) &= -\frac{3 z^2 g e^{-2A}}{L^2} \left[ A'' + A'\left(3A' -
\frac{6}{z} + \frac{3g'}{2g}\right) -
\frac{1}{z}\left(-\frac{4}{z} + \frac{3g'}{2g}\right) +
\frac{g''}{6g} \right],
\end{align}
with the determinant $\det\mathcal{G}$ defined via the following
$2\times2$ integral matrix:
\begin{equation}
\det \mathcal{G} =
\begin{vmatrix}
\displaystyle\int_{0}^{z_t} dy\, y^3 e^{-3A(y)} & \displaystyle\int_{0}^{z_t} dy\, y^3 e^{-3A(y)-c y^2} \\
\displaystyle\int_{z_t}^{z} dy\, y^3 e^{-3A(y)} &
\displaystyle\int_{z_t}^{z} dy\, y^3 e^{-3A(y)-c y^2}
\end{vmatrix}.
\end{equation}

The Hawking temperature of the black brane follows from bulk
surface gravity:
\begin{align}
T &= \frac{z_t^3 e^{-3A(z_t)}}{4\pi \displaystyle\int_{0}^{z_t}
dy\, y^3 e^{-3A(y)}} \left[ 1 + \frac{2c\mu^2 e^k \left( e^{-c
z_t^2} \displaystyle\int_{0}^{z_t} dy\, y^3 e^{-3A(y)} -
\displaystyle\int_{0}^{z_t} dy\, y^3 e^{-3A(y)-c y^2} \right)}{(1
- e^{-c z_t^2})^2} \right].
\end{align}

Within this machine learning calibrated EMD framework, the warp
factor ansatz is fixed as $A(z)=d\ln(az^2+1)+d\ln(bz^4+1)$, and
the dilaton dependent gauge kinetic function reads
$f(z)=e^{cz^2-A(z)+k}$. Six free coefficients $\{a,b,c,d,k,G_5\}$
are separately optimized for $N_f=0$, $N_f=2$, and $N_f=2+1$
through gradient descent machine learning, using lattice QCD
equations of state and baryon susceptibility as training targets.
All fitted constants and zero density crossover temperatures $T_c$
extracted from lattice matched thermodynamics are collected in
Table \ref{tab:emd_params}. For the $N_f=0$ pure gluon case, we
have $c=0$, which removes all baryon chemical potential dependence
from the metric functions, consistent with pure Yang-Mills theory
lacking quark degrees of freedom and baryonic charge.

\begin{table}[htbp]
  \centering
  \begin{tabular}{|c|c|c|c|c|c|c|c|}
    \hline
     & $a$ & $b$ & $c$ & $d$ & $k$ & $G_5$ & $T_c$ \\
    \hline
    $N_f = 0$ & $0$ & $0.072$ & $0$ & $-0.584$ & $0$ & $1.326$ & $0.265$ \\
    \hline
    $N_f = 2$ & $0.067$ & $0.023$ & $-0.377$ & $-0.382$ & $0$ & $0.885$ & $0.189$ \\
    \hline
    $N_f = 2 + 1$ & $0.204$ & $0.013$ & $-0.264$ & $-0.173$ & $-0.824$ & $0.400$ & $0.128$ \\
    \hline
  \end{tabular}
  \caption{Machine learning fitted bulk parameters from \cite{Chen:2024}, identical to the spacetime adopted for shooting string instantaneous energy loss in \cite{Zhang:2025}.
  Units: $G_5$ in $\text{GeV}^3$, $a,c$ in $\text{GeV}^2$, $b$ in $\text{GeV}^4$ and $T_c$ in $\text{GeV}$. $T_c$ denotes the zero density crossover temperature for each flavor plasma,
  obtained from lattice calibrated thermodynamic equations of state.}
  \label{tab:emd_params}
\end{table}

The three flavor configurations have distinct physical meanings.
The $N_f=0$ case corresponds to pure Yang-Mills theory, which
serves as a baseline for isolating the effects of dynamical
quarks. The $N_f=2$ configuration includes two light quark flavors
($u$ and $d$), approximating the light quark sector of QCD. The
$N_f=2+1$ configuration adds the strange quark, representing the
most realistic description of the QGP produced in heavy ion
collisions, where strangeness production plays an important role.
As validated in the original EMD machine learning publication and
the companion shooting string jet study, this six parameter setup
accurately reproduces lattice QCD equilibrium thermodynamics for
all three flavor sectors, including pressure, energy density, and
baryon susceptibility. It also yields distinct critical endpoint
temperature and chemical potential coordinates for $N_f=2$ and
$N_f=2+1$ QCD, reflecting the influence of dynamical quark flavors
on the phase diagram.

In what follows, we employ exactly this lattice fitted bulk
geometry to carry out falling string null geodesic calculations of
light quark thermalization stopping distance, enabling direct
quantitative comparison against published shooting string results
on the identical background.

\section{Light quark stopping distance via the falling string formalism}
\label{sec:fallingstring}

We now derive the light quark stopping distance using the WKB
semiclassical null geodesic scheme standard for falling string
holography. We first outline the holographic mapping between
boundary hard quarks and bulk gauge wave packets, derive the
differential relation connecting boundary longitudinal
displacement to radial coordinate $z$, integrate to obtain the
closed stopping distance integral, and construct the conformal
$\mathcal{N}=4$ SYM reference baseline for normalized ratio
analysis.

\subsection{Physical picture of the falling string configuration}

Under holographic duality, a high energy light quark propagating
through the four dimensional boundary QGP corresponds to a
massless U(1) gauge wave packet localized near the AdS boundary
$z\to0$ within the five dimensional bulk. The quark's large
boundary energy and three momentum translate into conserved bulk
momentum components, guaranteed by metric translational invariance
along all boundary spacetime directions. As the quark advances on
the boundary, its dual string endpoint sinks radially inward
toward the black brane horizon $z=z_t$. Once the endpoint reaches
$z_t$, the whole string is fully absorbed by the black brane, and
the high $p_T$ boundary jet completely thermalizes, losing all
coherent energy and momentum information. The total longitudinal
distance traveled by the light quark from injection until full
thermalization defines the thermalization stopping distance
$x_{\mathrm{stop}}$, an integrated observable summarizing total
cumulative medium induced energy loss over the entire propagation
path.

This setup differs fundamentally from the shooting string
framework, which describes quarks generated near the horizon
moving outward toward the boundary and extracts local differential
energy loss rates $dE/dx$ at individual points along the
trajectory. The two string formalisms therefore provide
complementary global and local probes of parton-medium coupling,
and their mutual consistency provides a stringent test of the
underlying holographic description.

\subsection{WKB approximation and null geodesic constraint}

For highly energetic light quarks carrying large momentum, the WKB
semiclassical approximation applies. The massless bulk gauge field
wavefunction factorizes into a rapidly oscillating plane wave
phase and slowly varying spatial envelope:
\begin{equation}
A_j(t,z) = \exp\left[\frac{i}{\hbar}\left(q_k x_k + \int dz\,
q_z\right)\right] \tilde{A}_j(t,z),
\end{equation}
where $q_k$ denote conserved boundary four momentum components,
$q_z$ momentum along the radial $z$ direction, and
$\tilde{A}_j(t,z)$ the slowly varying gauge field amplitude. In
the classical limit $\hbar\to0$, quantum interference disappears,
and massless bulk field equations reduce to the null geodesic
constraint $ds^2=0$ for the five dimensional line element, as
massless particles follow null trajectories in curved spacetime.
We adopt the string frame, where the metric is related to the
Einstein frame via $A_s(z) = A(z) + \sqrt{1/6}\,\phi(z)$. This
frame proves convenient for falling string descriptions, as string
worldsheet dynamics are naturally formulated using the string
frame metric:
\begin{equation}
ds^2 = \frac{L^2 e^{2A_s(z)}}{z^2} \left[ -g(z) dt^2 + d\vec{x}^2
+ \frac{dz^2}{g(z)} \right]. \label{eq:string frame}
\end{equation}

Substituting metric \eqref{eq:string frame} into $ds^2=0$ and
restricting to quarks propagating purely along the longitudinal
$x$ direction:
\begin{equation}
0 = g_{tt}dt^2 + g_{xx}dx^2 + g_{zz}dz^2.
\end{equation}
Parametrizing the full null trajectory using a monotonic affine
parameter $\zeta$ and rearranging to isolate radial velocity
gives:
\begin{equation}
\frac{dz}{d\zeta} = \frac{1}{\sqrt{g_{zz}}}
\left[-g_{tt}\left(\frac{dt}{d\zeta}\right)^2 -
g_{xx}\left(\frac{dx}{d\zeta}\right)^2 \right]^{1/2}.
\end{equation}

Metric translational symmetry implies covariant momenta $p_i =
g_{ij}dx^j/d\zeta$ conjugate to boundary coordinates are conserved
along the geodesic and proportional to physical quark four
momentum $q_i$ measured on the boundary QGP. Dividing the boundary
longitudinal velocity relation by radial velocity eliminates the
auxiliary affine parameter $\zeta$, yielding a direct differential
relation between boundary displacement $dx$ and radial coordinate
$dz$:
\begin{equation}
\frac{dx}{dz} = \sqrt{g_{zz}} \frac{g^{xx} q_x}{\left(-g_{tt}
q_t^2 - g_{xx} q_x^2\right)^{1/2}}.
\end{equation}

For light quarks moving along the $x$ axis, conserved boundary
four momentum simplifies to $q_i=(-\omega,0,0,|\vec{q}|)$, with
$\omega$ the quark total energy and $|\vec{q}|$ spatial momentum
magnitude. Substituting diagonal metric components from Eq.
\eqref{eq:string frame} cancels the universal conformal prefactor
$L^2 e^{2A_s(z)}/z^2$ across numerator and denominator, a generic
property of null geodesics under arbitrary bulk Weyl rescalings of
spacetime.

\subsection{Stopping distance integral and SYM reference normalization}

Integrating the differential displacement relation from the AdS
boundary $z=0$ to the black brane horizon $z=z_t$ produces the
closed integral for light quark thermalization stopping distance:
\begin{equation}
x_{\mathrm{stop}} = \int_{0}^{z_t}
\frac{dz}{\sqrt{\omega^2/|\vec{q}|^2 - g(z) }},
\label{eq:stopintegral}
\end{equation}
where $g(z)$ is the EMD blackening factor defined in Eq.
\eqref{blackening factor}. Physically, each infinitesimal radial
shift $dz$ of the string endpoint corresponds to a small
longitudinal propagation step $dx$ of the boundary quark;
integration over the full radial range accumulates the maximal
penetration depth before full thermalization. Larger $g(z)$ at
fixed $z$ reduces the square root denominator, increasing $dx/dz$
and extending total stopping distance, corresponding to weaker
medium energy loss and milder jet quenching.

To isolate nonconformal corrections originating from dynamical
quarks, finite temperature, and baryon density, we normalize all
computed stopping distances against a reference value
$x_{\text{SYM}}$ obtained from pure conformal $\mathcal{N}=4$ SYM
plasma without dynamical quarks. At high temperature where $z_t$
becomes small (horizon positioned close to the AdS boundary), the
EMD geometry reduces to standard AdS-Schwarzschild black branes,
yielding simplifications for the SYM baseline:
\begin{equation}
A_s(z) \approx 0,\quad g(z) \approx 1 - \frac{z^4}{z_t^4},\quad T
\approx \frac{1}{\pi z_t}.
\end{equation}

Inserting these expressions into Eq. \eqref{eq:stopintegral} gives
the conformal reference penetration depth $x_{\text{SYM}}$. All
numerical plots adopt the dimensionless ratio $x/x_{\text{SYM}}$
as the primary observable to quantify deviations from conformal
SYM dynamics, removing trivial overall scalings dependent on quark
energy $\omega$ and AdS length $L$. Within this ratio,
$x/x_{\text{SYM}}<1$ signifies shorter stopping distance and
stronger jet quenching, while $x/x_{\text{SYM}}>1$ corresponds to
weaker medium dissipation. This provides a direct benchmark
against the $|dE/dx|/|dE/dx|_{\text{SYM}}$ ratio reported in the
companion shooting string work \cite{Zhang:2025}.

\section{Numerical results and cross observable physical analysis}
\label{sec:results}

We perform high precision numerical integration of the stopping
distance integral Eq.~\eqref{eq:stopintegral} over broad
continuous ranges of reduced temperature $T/T_c$ and baryon
chemical potential $\mu$, separately evaluating the three flavor
parameter sets listed in Table~\ref{tab:emd_params}. Our numerical
analysis proceeds first with zero baryon density results, followed
by finite baryon chemical potential scans at fixed reduced
temperature.

The choice $|\vec{q}|=0.99\omega$ corresponds to a small
virtuality. Varying $|\vec{q}|/\omega$ between $0.95$ and $0.999$
changes $x/x_{\text{SYM}}$ by less than $1\%$, because for
$|\vec{q}|\approx\omega$ the integrand in
Eq.~\eqref{eq:stopintegral} is approximately $1-g(z)$. The results
are therefore robust against the precise choice of virtuality.

The RHIC beam energy scan program has explored the QCD phase
diagram up to $\mu \approx 420$ MeV in collider mode
\cite{STAR:2017}, and the fixed-target program has extended this
reach to $\mu \approx 720$ MeV \cite{Labonte:2025}. Lattice QCD
constraints on the equation of state are well controlled for
$\mu/T \lesssim 2$ through a Taylor expansion in $\mu/T$
\cite{Bazavov:2017}. At the temperatures considered in this work,
this corresponds to $\mu \lesssim 0.3$--$0.6$ GeV. We extend our
analysis to $\mu \leq 1.0$ GeV, which covers the experimentally
explored region and provides a qualitative extrapolation toward
the dense matter regime relevant for future facilities such as
FAIR and NICA. For $\mu/T \gtrsim 2$, the results should be
regarded as qualitative extrapolations beyond the region where
lattice QCD constraints on the equation of state are under
quantitative control. We do not apply the model to $\mu$ as large
as $3$ GeV, where additional phases such as color
superconductivity could appear.

\subsection{Zero baryon chemical potential: flavor ordering and crossover}

We first consider calculations carried out at vanishing baryon
chemical potential, where temperature acts as the sole varying
thermodynamic parameter.

\begin{figure}[htbp]
    \centering
    \includegraphics[width=0.5\linewidth]{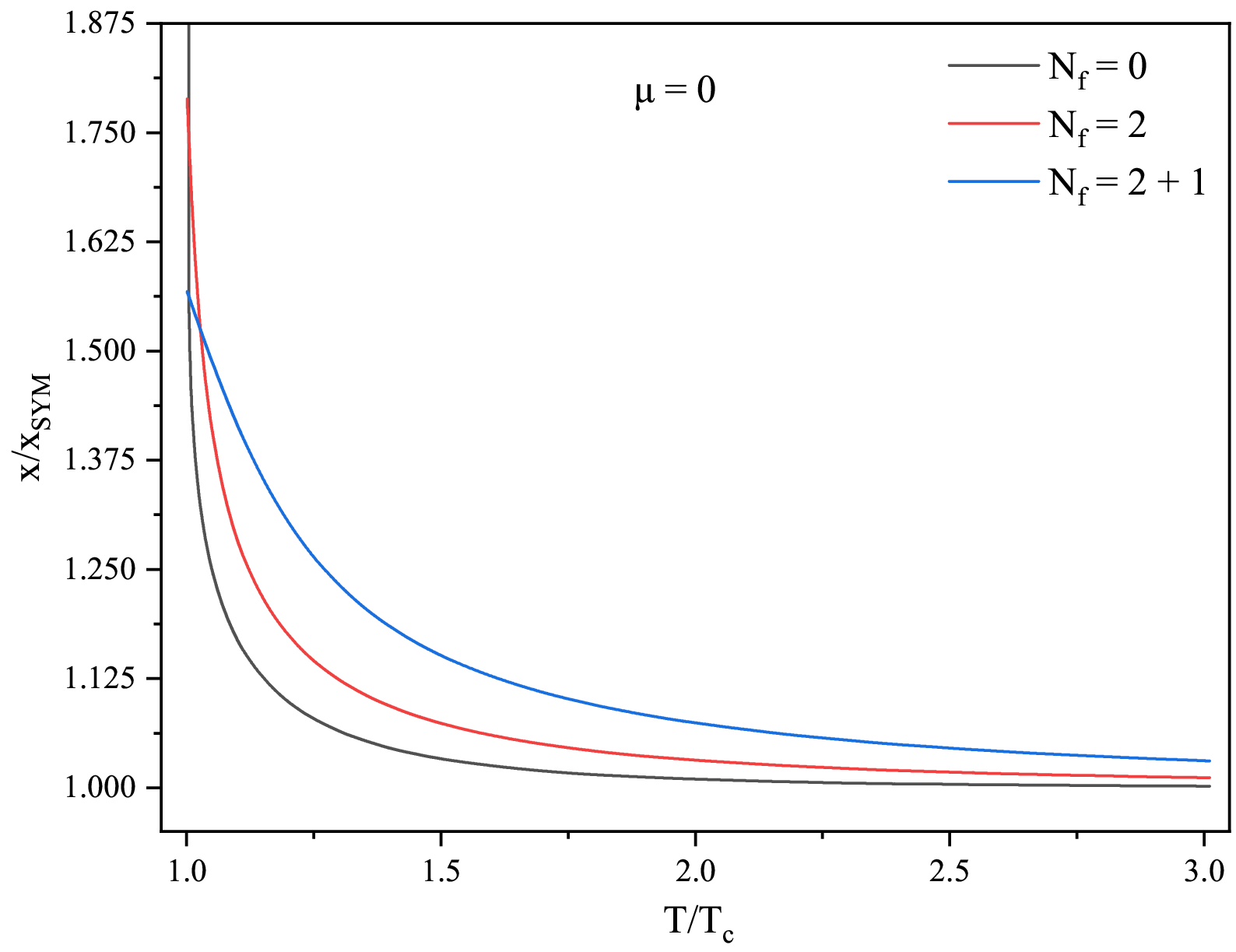}
    \caption{Normalized stopping distance ratio $x/x_{\text{SYM}}$ as a function of reduced temperature $T/T_c$ at vanishing baryon chemical potential $\mu=0$.
    Black solid line: $N_f=0$ pure gluon plasma; red solid line: $N_f=2$ two flavor QCD; blue solid line: $N_f=2+1$ QCD with strange quarks.
    The zero-density crossover temperatures are $T_c=0.265$ GeV ($N_f=0$), $T_c=0.189$ GeV ($N_f=2$), and $T_c=0.128$ GeV ($N_f=2+1$), as listed in Table~\ref{tab:emd_params}.
    Here we take $|\vec{q}|=0.99\omega$.}
    \label{fig:T/T_cscan_mu0}
\end{figure}

Figure \ref{fig:T/T_cscan_mu0} presents $x/x_{\text{SYM}}$ plotted
against reduced temperature $T/T_c$ at $\mu=0$ for all three quark
flavor systems. Two main features stand out. First,
$x/x_{\text{SYM}}$ decreases monotonically with rising $T/T_c$ and
converges to unity at asymptotically high reduced temperatures far
above each system's crossover temperature $T_c$. This convergence
is a geometric statement: as $T/T_c$ increases, the horizon $z_t$
shrinks and the EMD metric reduces to the AdS-Schwarzschild form
used for the SYM baseline. It does not imply that high-temperature
QCD literally becomes $\mathcal{N}=4$ SYM; rather, the
nonconformal corrections encoded in $A(z)$ become negligible over
the relevant integration range. This high $T/T_c$ convergence
matches lattice QCD thermodynamics, where differences between
$N_f=0$, $N_f=2$, and $N_f=2+1$ pressure and energy density vanish
for $T \gg T_c$. The ratio lies above unity at finite $T/T_c$
because the nonconformal corrections encoded in the EMD warp
factor $A(z)$ increase the effective stopping distance relative to
the conformal SYM baseline. As $T/T_c$ grows, these corrections
diminish and $x/x_{\text{SYM}}$ approaches unity from above.

The second feature is a nontrivial flavor ordering that evolves
with reduced temperature. Near $T/T_c=1$ the behavior is more
complex due to the interplay between the $b$- and $a$-dependent
terms in $A(z)$; for $T/T_c\gtrsim1.03$, however, the ratio
$x/x_{\rm SYM}$ follows the ordering $N_f=2+1>N_f=2>N_f=0$, which
corresponds to the quenching hierarchy $N_f=0>N_f=2>N_f=2+1$. This
quenching hierarchy agrees with that obtained from the local
energy loss rate $dE/dx$ in the companion shooting string
study~\cite{Zhang:2025} when the latter is re-expressed in the
reduced-temperature framework. The two independent holographic
observables, one integrated and one local, therefore agree on the
flavor dependence of jet quenching away from the immediate
vicinity of the crossover, supporting the robustness of the
flavor-resolved EMD description.

\subsection{Finite baryon chemical potential: density driven quenching enhancement}

We next turn to results with finite baryon chemical potential,
examining two distinct fixed reduced temperature values,
$T/T_c=1.10$ and $T/T_c=1.50$, to isolate temperature-dependent
density sensitivity.

\begin{figure}[htbp]
    \centering
    \includegraphics[width=7.8cm]{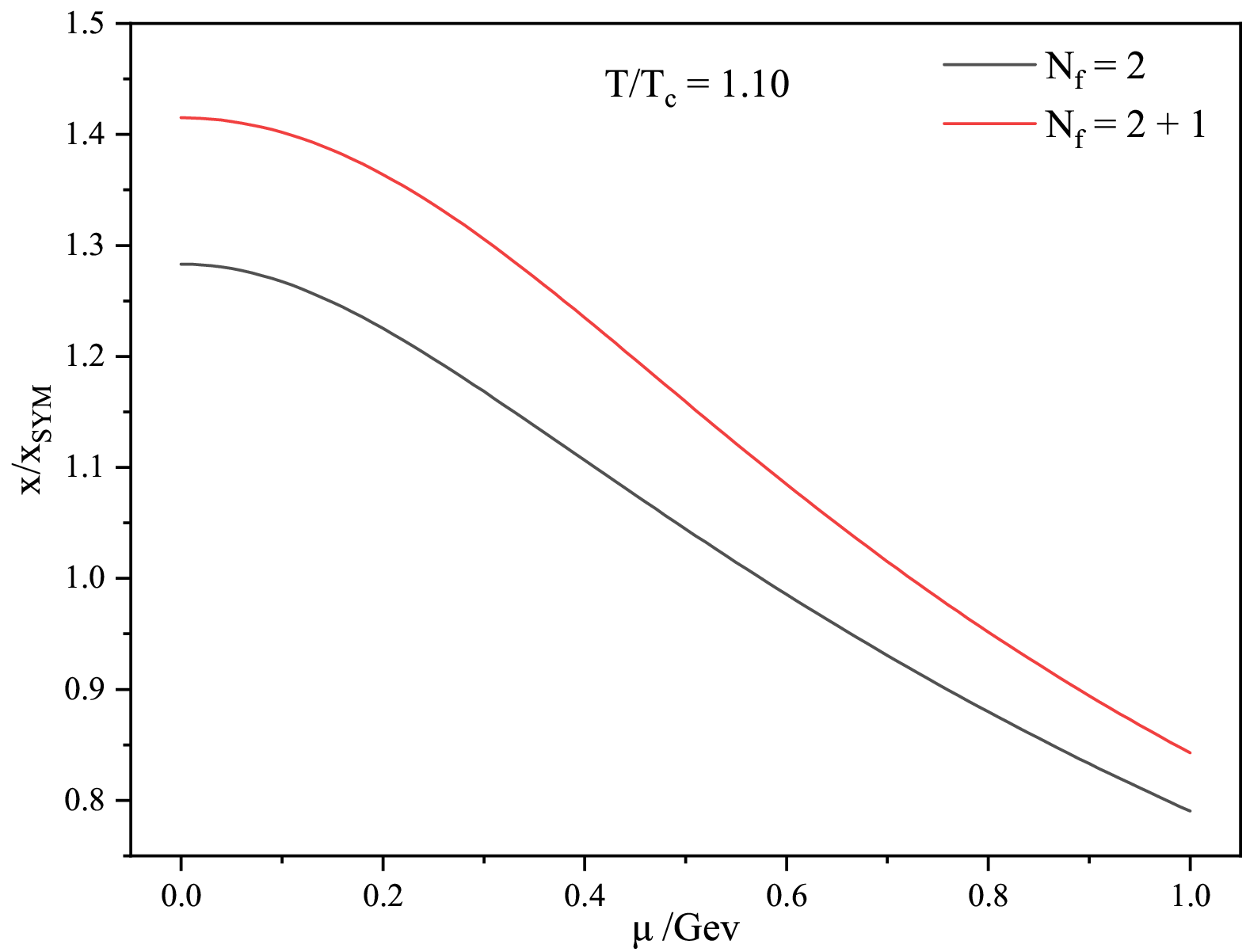}
    \includegraphics[width=7.8cm]{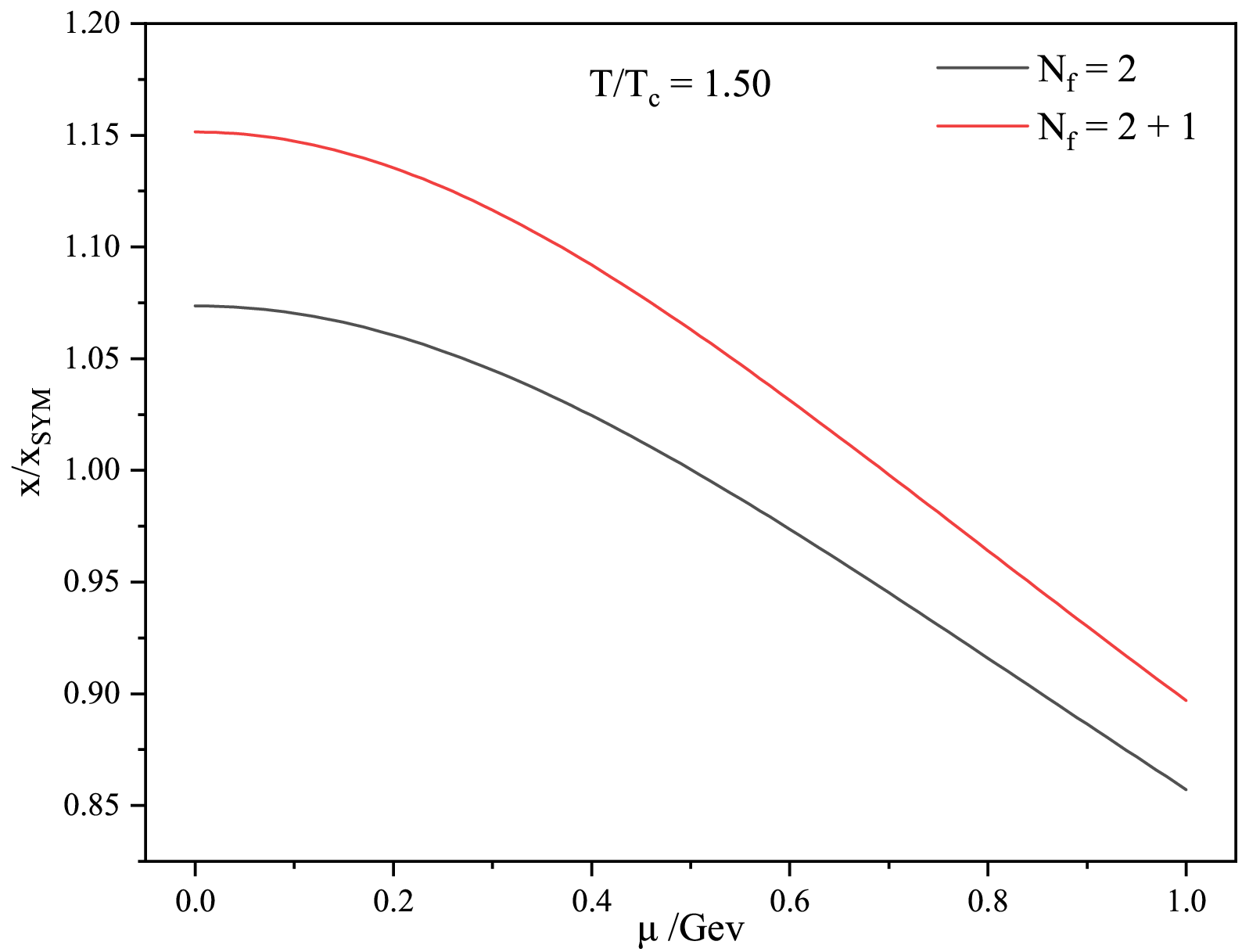}
    \caption{Normalized stopping distance ratio $x/x_{\text{SYM}}$ as a function of baryon chemical potential $\mu$ at fixed reduced temperature. Left panel: $T/T_c=1.10$; Right panel: $T/T_c=1.50$. Black curves: $N_f=2$ two flavor plasma; red curves: $N_f=2+1$ QCD with strange quarks. The pure gluon $N_f=0$ geometry has no baryon density dependence and is omitted here. Here we take $|\vec{q}|=0.99\omega$. The $T_c$ values are the same as in Fig.~\ref{fig:T/T_cscan_mu0}.}
    \label{fig:mu_scan_fixedT}
\end{figure}

For both flavor configurations, the stopping distance ratio falls
monotonically with increasing $\mu$ at constant $T/T_c$. This
demonstrates that higher equilibrium baryon number density
universally shortens light quark thermalization penetration depth
and amplifies total medium induced energy loss. The $N_f=2+1$
system consistently yields a larger stopping distance ratio than
$N_f=2$ at both reduced temperatures, corresponding to weaker
integrated quenching, consistent with the flavor hierarchy
observed in Fig.~\ref{fig:T/T_cscan_mu0}. This density-driven
enhancement of quenching is not specific to the falling string
formalism. Within the same EMD background, consistent trends have
been observed across falling-string stopping distances,
shooting-string instantaneous energy loss, holographic drag
coefficients, and jet-quenching parameter calculations, ruling out
formalism-dependent bias in finite-density predictions.

Furthermore, in the AdS-Reissner-Nordstr\"om (AdS-RN) and
AdS-RN-Vaidya backgrounds, the stopping distance of a falling
string was found to decrease monotonically with increasing
chemical potential, corresponding to stronger quenching in denser
media \cite{Caceres:2014}. Within the EMD framework, the
jet-quenching parameter $\hat{q}$, the heavy-quark drag force, and
the Langevin diffusion coefficients are all enhanced with
increasing baryon density \cite{Grefa:2022,Rougemont:2016}. These
consistent findings across different string configurations and
geometric setups indicate that density-driven enhancement of
quenching is a robust feature of strongly coupled holographic
plasmas, rather than an artifact of any specific formalism.

Comparing the two reduced-temperature panels reveals that baryon
chemical potential exerts a drastically stronger quenching effect
at lower reduced temperature. At $T/T_c=1.10$, $x/x_{\text{SYM}}$
declines steeply with rising $\mu$, while the suppression slope
weakens substantially at elevated $T/T_c=1.50$. This
reduced-temperature dependent density sensitivity originates from
proximity to the QCD critical endpoint: at lower $T/T_c$, the
plasma lies much closer to the CEP on its flavor phase diagram,
where medium transport susceptibilities and coupling strength
become extremely sensitive to small baryon density perturbations.
This behavior is consistent with the expectation that medium
transport susceptibilities and coupling strength vary most rapidly
in the vicinity of the critical endpoint.

This holographic result provides a qualitative connection to the
primary scientific goal of the RHIC beam energy scan program.
Experimental searches for enhanced fluctuations of jet observables
such as $R_{AA}$ as collision energy (and hence medium baryon
chemical potential) varies are motivated by the expectation that
medium properties change rapidly near the QCD critical endpoint.
Our finding that the quenching strength varies sharply under small
$\mu$ shifts near the CEP is consistent with this expectation.
However, the stopping distance is an integrated quantity and
cannot be directly equated with $R_{AA}$; a quantitative link
would require a dedicated transport analysis, as performed in the
phenomenological studies of Refs.~\cite{ss2,Ficnar:2011}. We
therefore regard this as a qualitative consistency check.

\subsection{Joint temperature and chemical potential dependence}

Finally, we examine joint variations of reduced temperature and
baryon chemical potential, with separate plots for the $N_f=2$ and
$N_f=2+1$ systems.

\begin{figure}[htbp]
    \centering
    \includegraphics[width=7.8cm]{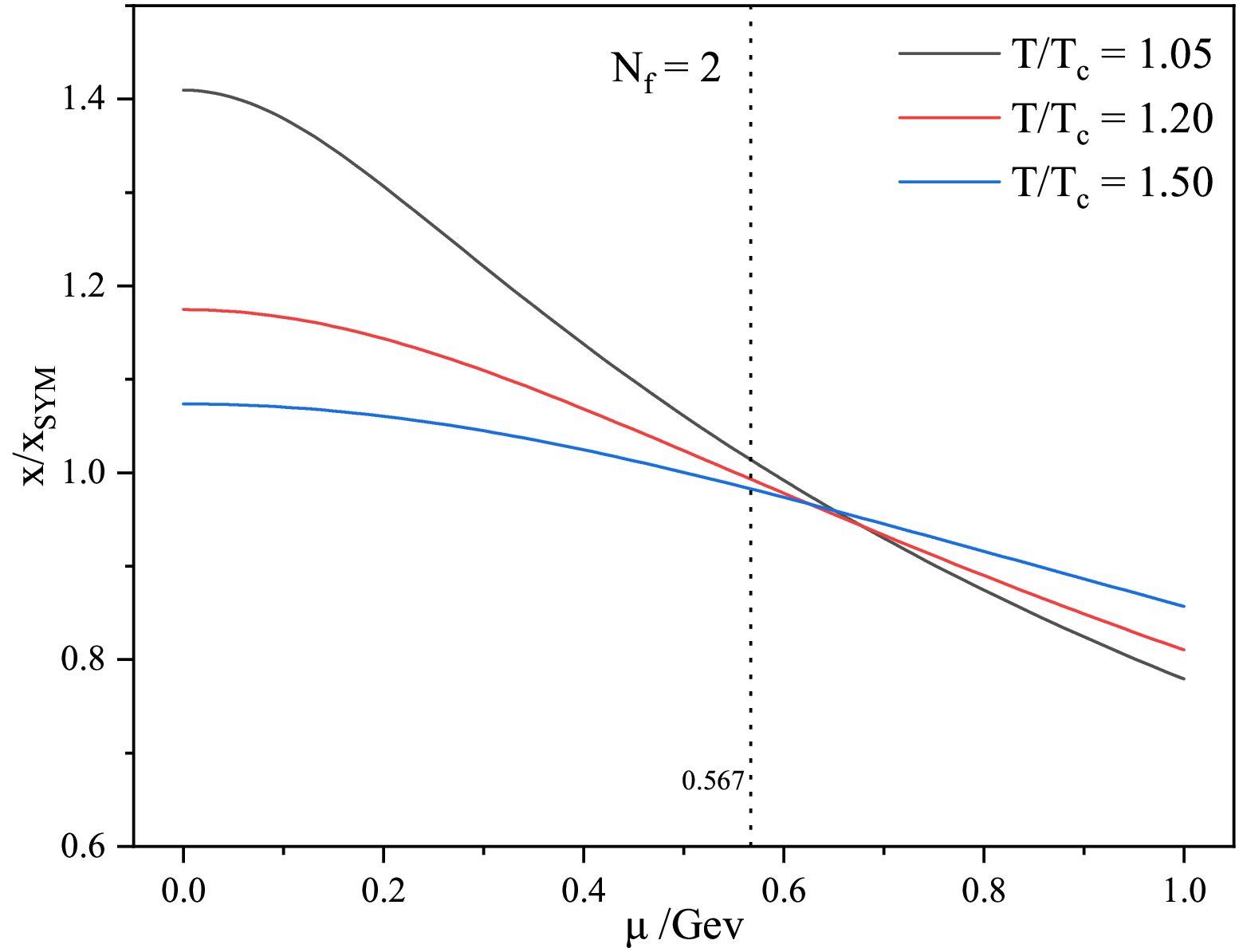}
    \includegraphics[width=7.8cm]{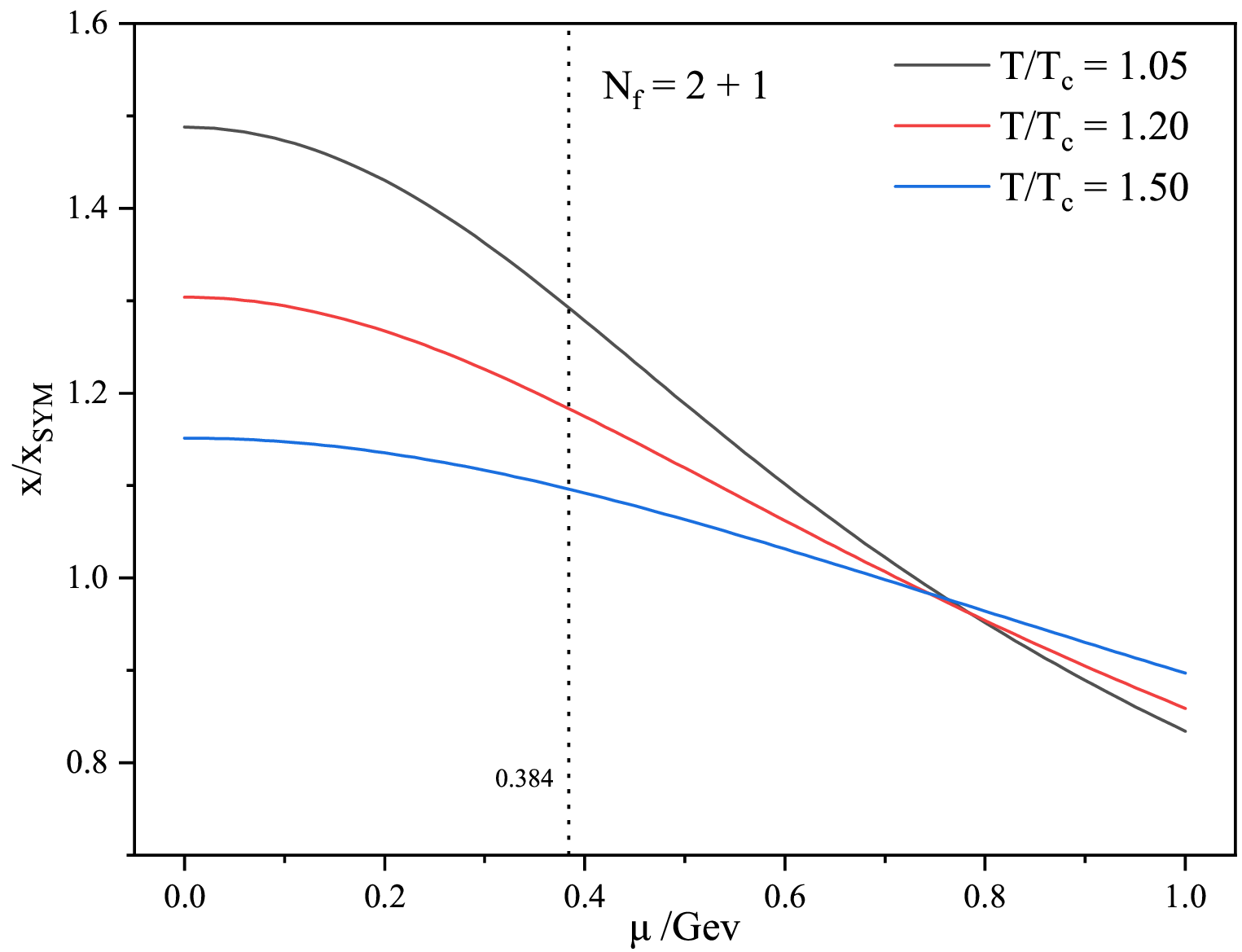}
    \caption{Normalized stopping distance ratio $x/x_{\text{SYM}}$
versus baryon chemical potential $\mu$ for three fixed reduced
temperatures. Left panel: $N_f=2$ two flavor plasma; Right panel:
$N_f=2+1$ QCD with strange quarks. Black, red, and blue curves
correspond to $T/T_c=1.05$, $1.20$, and $1.50$, respectively. The
vertical dotted line marks the $\mu/T=2$ boundary for the
$T/T_c=1.50$ curve, which is the most conservative among the three
curves shown. To the right of this line, all results should be
regarded as extrapolations beyond the region where lattice QCD
constraints on the equation of state are under quantitative
control \cite{Bazavov:2017}. Here we take $|\vec{q}|=0.99\omega$.
The $T_c$ values are the same as in Fig.~\ref{fig:T/T_cscan_mu0}.}
\label{fig:mu_T_joint_scan}
\end{figure}

Figure \ref{fig:mu_T_joint_scan} isolates the combined dependence
on reduced temperature and baryon chemical potential by plotting
$x/x_{\text{SYM}}$ against $\mu$ at three distinct reduced
temperatures for the $N_f=2$ and $N_f=2+1$ systems. For both
flavor sectors, the stopping distance ratio decreases
monotonically with increasing $\mu$ at fixed $T/T_c$, confirming
that finite baryon density universally enhances integrated
quenching. Comparing the two panels, the overall magnitude of
$x/x_{\text{SYM}}$ is systematically larger for $N_f=2+1$ than for
$N_f=2$, consistent with the weaker integrated quenching of the
strange-quark medium found at high $T/T_c$ in
Fig.~\ref{fig:T/T_cscan_mu0}.

A notable feature in both panels is the crossing of the three
curves at $\mu \approx 0.6$--$0.7$ GeV. At low chemical potential,
lower reduced temperatures yield larger stopping distance ratios,
i.e., weaker quenching. As $\mu$ increases, this ordering
reverses: at $\mu \gtrsim 0.7$ GeV, the lowest reduced temperature
$T/T_c=1.05$ produces the smallest ratio and hence the strongest
quenching. This reversal can be understood as follows. Within a
fixed flavor sector, a lower $T/T_c$ corresponds to a lower
absolute temperature, placing the medium closer to the critical
region. Near-critical media respond more strongly to baryon
density perturbations, so the $T/T_c=1.05$ curve falls more
steeply with increasing $\mu$ than the higher-$T/T_c$ curves. Once
$\mu$ becomes sufficiently large, this enhanced density
sensitivity overtakes the initial offset in $x/x_{\text{SYM}}$ at
$\mu=0$, and the $T/T_c=1.05$ curve becomes the lowest of the
three. This behavior is consistent with the expectation that
medium transport susceptibilities and coupling strength vary most
rapidly in the vicinity of the critical endpoint.

We emphasize that the crossing behavior described above occurs at
$\mu \approx 0.6$--$0.7$ GeV, which lies entirely in the
extrapolation region for all three curves. The vertical dotted
line in each panel marks the $\mu/T=2$ boundary for the
$T/T_c=1.50$ curve, which is the most conservative among the
three; for the $T/T_c=1.05$ and $1.20$ curves, the corresponding
boundaries lie at smaller $\mu$. The crossing should therefore be
regarded as a qualitative prediction of the EMD model rather than
a quantitative result.

Finally, a side-by-side comparison of the two flavor sectors
reveals a characteristic flavor-dependent feature. Systems
containing more quark flavors possess a lower critical
temperature, which shifts the physically accessible near-critical
region toward smaller absolute $T$. This agrees with
well-established lattice QCD results stating that the
deconfinement transition temperature decreases as the number of
light quark flavors increases. At similar $T/T_c$, the magnitude
of the zero-$\mu$ deviation from SYM is smaller for the $N_f=2+1$
system, indicating that nonconformal deviations near the crossover
weaken as flavor content grows within this holographic framework.
Nevertheless, the qualitative evolution with chemical potential
remains identical for both systems: raising $\mu$ always enhances
energy loss, and the sensitivity peaks near the phase transition.
This universality suggests that chemical-potential-driven
enhancement of jet quenching is a robust feature of strongly
coupled QCD-like matter within the holographic EMD description.

\section{Discussion}
\label{sec:discussion}

The flavor dependence of jet quenching in our flavor-resolved EMD
holographic framework exhibits a nontrivial evolution with reduced
temperature. In the reduced-temperature plane, the three systems
have different absolute temperatures because their $T_c$ values
differ, so the observed ordering reflects a combination of the
intrinsic metric parameters and this normalization effect.

It is important to note that the apparent weakening of quenching
with increasing $N_f$ at fixed $T/T_c$ does not contradict the
intuition that a medium with more degrees of freedom has larger
energy density. At fixed \emph{absolute} temperature, the
companion shooting-string study finds that $|dE/dx|$ increases
with $N_f$, i.e., quenching is enhanced. The reversal in the
$T/T_c$ framework arises because systems with more flavors have
lower $T_c$, so at the same $T/T_c$ they are compared at different
absolute temperatures. The integrated stopping distance is
sensitive to the full radial integration range; the larger $a$
parameter for $N_f=2+1$ shifts the horizon outward, lengthening
the path. Thus the flavor ordering in the reduced-temperature
plane is a combined effect of intrinsic metric parameters and the
normalization by different $T_c$ values, rather than a genuine
weakening of quenching at fixed absolute temperature.

For $T/T_c\gtrsim1.03$, the ratio $x/x_{\text{SYM}}$ ordering
stabilizes to $N_f=2+1>N_f=2>N_f=0$, corresponding to the
quenching hierarchy $N_f=0>N_f=2>N_f=2+1$. In this regime, the
horizon coordinate $z_t$ is small and the total radial integration
range plays a dominant role. Systems with more dynamical quark
flavors have larger fitted $a$ parameters; the $a$ coefficient
modulates the $z^2$ term in $A(z)$, shifting $z_t$ outward at
fixed $T/T_c$. A larger $z_t$ lengthens the integration range in
Eq.~\eqref{eq:stopintegral}, increasing the total stopping
distance and weakening jet quenching. This provides a consistent
geometric explanation for the high-$T/T_c$ ordering.

For $1.000<T/T_c<1.03$, the ratio $x/x_{\text{SYM}}$ exhibits a
multi-stage evolution with two rapid changes. In the immediate
vicinity of $T/T_c=1$, the ordering is $N_f=0>N_f=2>N_f=2+1$, with
$N_f=0$ showing the weakest quenching. As $T/T_c$ increases
slightly, the ordering shifts to $N_f=2>N_f=0>N_f=2+1$, with
$N_f=2$ now the weakest quencher. These changes reflect the
changing relative importance of the $a$ and $b$ terms in $A(z)$
together with the shift in absolute temperature associated with
the reduced-temperature normalization. A simple one-parameter
explanation of this multi-stage evolution is not available; we
present it as a prediction of the full numerical solution.

It is instructive to compare the high-$T/T_c$ ordering with the
companion shooting string study~\cite{Zhang:2025}. There, $dE/dx$
was reported at fixed absolute temperature with the ordering
$|dE/dx|_{N_f=2+1}>|dE/dx|_{N_f=2}>|dE/dx|_{N_f=0}$. When
re-expressed in the reduced-temperature framework, this becomes
$N_f=0>N_f=2>N_f=2+1$, in agreement with the falling-string result
for $T/T_c\gtrsim1.03$. This consistency across two independent
string observables supports the robustness of the flavor-resolved
EMD description at high reduced temperature.

At low $T/T_c$, however, the two observables exhibit different
flavor orderings. This difference is not a contradiction but a
reflection of their complementary sensitivity to the bulk
geometry: the local quantity $dE/dx$ depends only on the metric at
the string endpoint, while the integrated stopping distance is
sensitive to the horizon position and the full radial integration
range. This distinction explains why the flavor ordering of the
two observables need not coincide near the crossover, while they
converge at higher $T/T_c$.

Finally, we note that the weakening of flavor sensitivity at high
$T/T_c$, where the three curves approach one another, suggests
that for jets probing the highest temperature regions of the
plasma the flavor composition of the medium becomes less
important. Whether this trend persists in the absolute-temperature
comparison relevant for LHC measurements requires further study.

\section{Conclusion and outlook}
\label{sec:conclusion}

In this work, we have carried out a systematic holographic
investigation of light quark thermalization stopping distances
using the null geodesic WKB formalism for falling strings on a
machine-learning-calibrated EMD background matching the spacetime
geometry used in recent shooting string simulations. Bulk metric
parameters are constrained to lattice QCD equations of state and
baryon susceptibility data for three flavor configurations:
$N_f=0$, $N_f=2$, and $N_f=2+1$. We have reviewed the derivation
of the stopping distance integral following \cite{ss,pm,pa} and
performed extensive scans over reduced temperature $T/T_c$ and
baryon chemical potential $\mu$. For $T/T_c\gtrsim1.03$, the
falling-string stopping distance and the shooting-string $dE/dx$
yield a consistent quenching hierarchy $N_f=0>N_f=2>N_f=2+1$ when
both are expressed in the reduced-temperature framework. Two rapid
transitions occur in the narrow window $1.00<T/T_c<1.03$,
reflecting the combined effect of the $a$ and $b$ terms in $A(z)$
together with the different $T_c$ values of the three systems.
This agreement at high $T/T_c$ provides a nontrivial cross-check
of the flavor-resolved EMD description, while the low-$T/T_c$
difference between local and integrated observables reflects their
complementary sensitivity to the bulk geometry.

The framework offers improved phenomenological applicability for
interpreting RHIC and LHC data: it modulates medium coupling
through quark flavor content, temperature, and baryon density.
Higher baryon density monotonically strengthens energy loss,
corresponding to stronger high-$p_T$ suppression in denser
fireballs. Collision systems near the QCD critical endpoint are
expected to show enhanced sensitivity to small baryon density
fluctuations, consistent with anomalous signals in RHIC beam
energy scan data, and dynamical strange quarks modify the
quenching magnitude in a manner consistent with measured strange
hadron yields.

The stopping distance integral can in principle be incorporated
into existing parton transport models, complementing the widely
used shooting string $dE/dx$ module. A combined analysis based on
both local and integrated energy loss observables could help
reduce systematic uncertainties in future data-model comparisons.
Flavor-resolved stopping distances can be evaluated pointwise on
hydrodynamic grid cells, allowing step-by-step tracking of parton
energy loss along realistic propagation trajectories. Future
directions include integrating the framework into event-by-event
$3+1$ dimensional hydrodynamic simulations, extending the machine
learning calibration to dynamical charm quarks, incorporating
transient magnetic fields by extending the EMD framework, and
adopting unsupervised machine learning to reconstruct the warp
factor $A(z)$ without preset functional forms.

\section*{Acknowledgments}
This work is supported by the National Natural Science Foundation
of China (NSFC) under grant No. 12375140 and the Fundamental
Research Funds for National Universities, China University of
Geosciences No. 2025XLB102.

\end{document}